\documentclass{aa}  
\usepackage{graphicx}
\usepackage{txfonts}
\RequirePackage{silence}
\usepackage{tabularx}

\usepackage[pdfpagelabels=false]{hyperref} 
\hypersetup{colorlinks=true,linkcolor=black,citecolor=blue,filecolor=blue,urlcolor=blue}
\begin{document}

   \title{The atomic C/O ratio of KELT-9b}

   \author{
                S.~Pelletier\inst{1}\thanks{Corresponding author: \texttt{Stefan.Pelletier@unige.ch}},
                N.~W.~Borsato\inst{2,3,4},
                J.~L.~Bean\inst{5},
                A.~Seifahrt\inst{6},
                D.~Kasper\inst{5},
                J.~Stürmer\inst{7},
                A.~R.~Costa Silva\inst{8,9,1},
                D.~Ehrenreich\inst{1,10},
                L.~Fossati\inst{11},
                H.~J.~Hoeijmakers\inst{4},
                B.~Prinoth\inst{12,4},
                M.~Steiner\inst{1},
                B.~Thorsbro\inst{13,4},
                V.~Vaulato\inst{1},
                D.~B.~Zucker\inst{3}
               }

       \institute{
                Observatoire astronomique de l'Université de Genève, 51 chemin Pegasi 1290 Versoix, Switzerland
             \and
                Univ Toulouse, CNES, CNRS, IRAP, 14 avenue Edouard Belin, 31400 Toulouse, France
             \and
                 School of Mathematical and Physical Sciences, Macquarie University, Sydney, NSW 2109, Australia
             \and
                Lund Observatory, Division of Astrophysics, Department of Physics, Lund University, Box 118, 221 00 Lund, Sweden
            \and
                Department of Astronomy \& Astrophysics, University of Chicago, Chicago, IL 60637, USA
            \and
                Gemini Observatory/NSF NOIRLab, 670 N. A'ohoku Place, Hilo, HI 96720, USA
            \and
                Landessternwarte, Zentrum f{\"{u}}r Astronomie der Universität Heidelberg, K{\"{o}}nigstuhl 12, D-69117 Heidelberg, Germany
            \and
                Instituto de Astrof\'{i}sica e Ci\^{e}ccias do Espaço, Universidade do Porto, CAUP, Rua das Estrelas, 4150-762 Porto, Portugal
            \and
                Departamento de F\'{i}sica e Astronomia, Faculdade de Ci\^{e}ncias, Universidade do Porto, Rua do Campo Alegre, 4169-007 Porto, Portugal
            \and
                Centre Vie dans l’Univers, Facult\'{e} des sciences de l’Universit\'{e} de  Gen`{e}ve, Quai Ernest-Ansermet 30, 1205 Geneva, Switzerland
            \and
                Space Research Institute, Austrian Academy of Sciences, Schmiedlstrasse 6, 8042 Graz, Austria
            \and
                European Southern Observatory, Karl-Schwarzschild-Strasse 2, 85748 Garching, Germany
            \and
                Université Côte d’Azur, Observatoire de la Côte d’Azur, CNRS, Laboratoire Lagrange, 06000 Nice, France
                 }

   \date{Received XXX; accepted XXX}

  \abstract
   {The carbon-to-oxygen (C/O) ratio of a giant planet's atmosphere has long been theorised to hold compositional information that can be traced back to its formation history.  Typically, the C/O ratio of an exoplanetary atmosphere is inferred from abundance measurements of major C- and O-bearing molecules such as CO, CO$_2$, H$_2$O, and OH. However, some exoplanets have such elevated temperatures that molecules can be nearly completely thermally dissociated at the pressure levels that observations probe, making it difficult to measure their C/O ratio via these traditional tracers.}
   {Here, rather than using molecules, we aim to retrieve the C/O ratio of KELT-9b (T$_\mathrm{eq} = 4000$\,K) directly from atomic C and O, which are detected in its atmosphere.}
   {We analysed two transits of KELT-9b observed with the MAROON-X high-resolution spectrograph, finding absorption cross-correlation signals from atomic C and O as well as refractory metals.  From this, we inferred the vertical temperature structure, the relative proportions of volatile and refractory species, and the C/O ratio of KELT-9b using a 1D local thermodynamic equilibrium atmospheric retrieval framework applied only to spectral regions mostly unaffected by non-local thermodynamic equilibrium effects.}
   {We measure the atomic C/O ratio of the atmosphere of KELT-9b to be $0.20_{-0.07}^{+0.13}$, which is slightly lower than the stellar value of $0.38\pm0.15$ and significantly below the solar value of $0.59\pm0.07$.  We otherwise confirm previous investigations of the terminator region, finding KELT-9b's atmosphere to be thermally inverted and slightly metal-rich. We measure the volatile-to-refractory ratio, a proxy for the ice-to-rock ratio, to be $1.24_{-0.78}^{+1.93}$ $\times$ solar ($[$M$_{\mathrm{vol}}$/M$_{\mathrm{ref}}]$ = $0.09_{-0.20}^{+0.19}$), which is consistent with both KELT-9 and the Sun.}
   {While there are caveats to interpreting observations of such an extreme planet, our results demonstrate that atomic species can be used to measure C/O ratios of exoplanetary atmospheres in high-temperature regimes where molecules are thermally dissociated.}


    \authorrunning{Pelletier et al.}
    \titlerunning{The C/O ratio of KELT-9b}

   \maketitle
   \nolinenumbers

\section{Introduction}
After hydrogen and helium, carbon and oxygen are the two most abundant elements in the Universe. In typical planetary and stellar atmospheres, they account for the majority ($\sim$72\%) of the total metal budget~\citep[][]{asplund_chemical_2021}. Importantly, C and O also have different enough condensation temperatures ($\sim$35\,K for C versus $\sim$155\,K for O at typical disc midplane pressures;~\citealt{minissale_thermal_2022}) that their relative gaseous and condensed proportions can vary with orbital separation in planet-forming regions of protoplanetary discs~\citep{oberg_effects_2011}. This has made the carbon-to-oxygen (C/O) ratio a high-interest observable in exoplanetary atmospheres, due to its potential as a tracer of planet formation~\citep[e.g.][]{madhusudhan_toward_2014, madhusudhan_atmospheric_2017, line_systematic_2014, line_solar_2021, barman_simultaneous_2015, lavie_heliosretrievalopen-source_2017, nowak_peering_2020, pelletier_where_2021, wang_retrieving_2022, brogi_roasting_2023, boucher_co_2023, bazinet_subsolar_2024, smith_roasting_2024, weiner_mansfield_metallicity_2024}.

\begin{figure*}[t]
    \includegraphics[width=\linewidth]{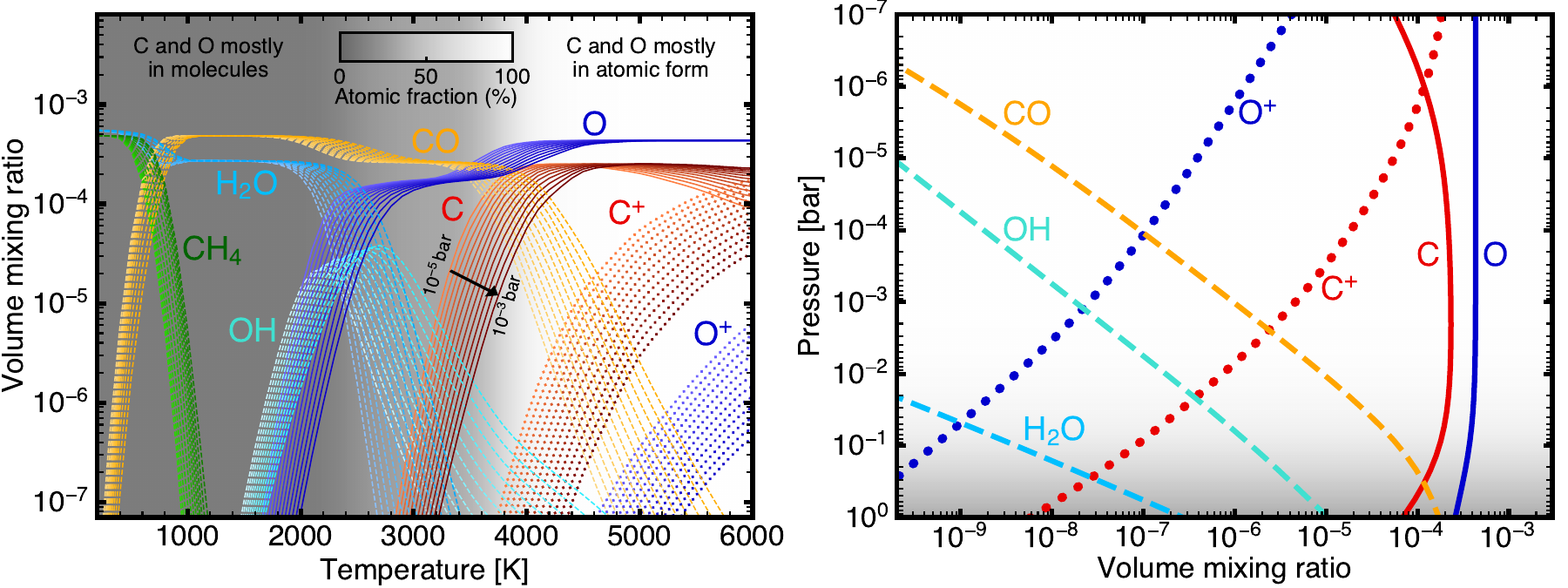}
    \centering
    \caption[]{
    Overview of the gas-phase oxygen and carbon chemistry in hydrogen-dominated planetary atmospheres. 
    \textit{Left}: Evolution of abundances as a function of temperature for important C- and O-bearing species predicted by chemical equilibrium for a solar composition at different pressures using \texttt{FastChem}. Solid lines show neutral elements, dotted lines ions, and dashed lines molecules.  Sets of coloured lines show a gradient between calculated abundances for different chemical species between 10$^{-5}$\,bar (lighter colour tone) and 10$^{-3}$\,bar (darker colour tone). The grey background shading represents the percentage of the C and O budget in atomic form (dark grey = 0\%, white = 100\%) at 10$^{-4}$\,bar. At low temperatures ($<$500\,K), most C and O atoms are held in H$_2$O and CH$_4$ molecules. Beyond $\sim$600\,K, CO replaces CH$_4$ as the main C carrier. Above $\sim$2000\,K, H$_2$O begins to thermally dissociate into atomic O and OH, with the latter becoming unstable and breaking apart into its atomic components past $\sim$2500\,K. Although particularly stable owing to its triple bond, even CO molecules dissociate beyond $\sim$3500\,K, which eventually results in the full C and O inventory being in atomic form. 
    \textit{Right}: Equilibrium chemistry abundance profiles assuming an isothermal temperature of 5000\,K. At such high temperatures, molecules are unstable in all but the deep atmosphere (shaded grey) and nearly all C and O is in neutral atomic form, with ionisation becoming important at microbar levels. While most known planetary atmospheres are in the molecule-dominated regime, extreme exoplanets such as KELT-9b have atomic-dominated atmospheres for which traditional means of determining the C/O ratio based on CO/H$_2$O/OH are no longer viable and directly measuring C and O abundances from their atomic forms is necessary. 
    }
    \label{fig:chemistry}
\end{figure*}

Accurately inferring the C/O ratio of a planetary atmosphere requires first identifying all relevant major C- and O-bearing species of the given temperature regime. In typical hot Jupiter atmospheres (T$_\mathrm{eq}$ of $\sim$ 1000 -- 2000\,K), C and O atoms are expected to be primarily bound in CO and H$_2$O molecules at the pressures probed by transmission spectroscopy (Fig.~\ref{fig:chemistry}, left panel). At lower temperatures (T$_\mathrm{eq}$ $\lesssim$ 900\,K), CH$_4$ can also be present~\citep{carleo_gaps_2022, bell_methane_2023, sing_warm_2024, welbanks_high_2024}, becoming more and more energetically favoured and eventually replacing CO as the primary C-bearing species in colder (T$_\mathrm{eq}$ $\lesssim$ 500\,K) hydrogen-dominated gaseous atmospheres, as is the case for Jupiter and Saturn~\citep{niemann_composition_1998, fletcher_methane_2009}. On the other hand, for hotter planets with atmospheric temperatures exceeding 2000\,K, OH and atomic O become important contributors to the total O budget as a result of the thermal dissociation of H$_2$O~\citep{parmentier_thermal_2018, brogi_roasting_2023, gandhi_revealing_2024}. Under different atmospheric conditions, other important carriers of C and O can also include CO$_2$ if the metallicity is elevated~\citep{lodders_atmospheric_2002}, or HCN and C$_2$H$_2$ in carbon-rich (C/O > 1) environments~\citep{madhusudhan_co_2012}.

Robustly measuring the C/O ratio of an exoplanetary atmosphere requires first obtaining abundance constraints of all major C- and O-bearing molecules. However, even in the scenario where all detectable atmospheric components are measured, some fraction of the total C and O budgets may still be missing from the observable gas phase. For example, the condensation of certain compounds (e.g. MgSiO$_3$, Mg$_2$SiO$_4$, and Fe$_2$O$_3$) can hold a significant amount of oxygen in liquid or solid form~\citep{burrows_chemical_1999, lodders_solar_2003}, which could result in biases if measuring the O abundance from only species in the gas phase. Estimating the exact amount of missing O can also be difficult as it depends on the availability of condensable refractory elements\footnote{Here `refractories' refer to species with relatively high ($\gtrsim 500$\,K) condensation temperatures (e.g. Fe, Mg, Ti, Na, Ca, Mn, and V). From a planet formation standpoint, refractories are generally accreted in the solid phase.}~\citep{turrini_tracing_2021, chachan_breaking_2023, fonte_oxygen_2023}, which are not necessarily accreted in similar proportions as volatile elements\footnote{In contrast to refractories, `volatiles' are species with relatively low condensation temperatures that tend to remain in the gas phase (e.g. C, N, and O).} during formation~\citep[e.g.][]{lothringer_new_2021, smith_roasting_2024, lothringer_refractory_2025, pelletier_crires_2025, sanchez_stellar_2026}. Even for some ultra-hot planets in synchronous rotation with dayside temperatures exceeding 2500\,K -- past which no significant cloud mass can be sustained -- condensation may still be important on the cooler permanent nightside~\citep[e.g.][]{ehrenreich_nightside_2020, hoeijmakers_hot_2020, pelletier_vanadium_2023,pelletier_breaking_2026,  pelletier_enriched_2026}, effectively sequestering a fraction of the O budget from the gas phase at the photospheric pressures probed by observations~\citep{parmentier_3d_2013, parmentier_transitions_2016, helling_sparkling_2019, helling_cloud_2021}. Measurements of the gas phase alone therefore do not necessarily reflect the bulk envelope composition, potentially resulting in an overestimation of the C/O ratio if not considering oxygen atoms preferentially held in condensates.

For even hotter exoplanets entering the stellar regime (T$_\mathrm{eq}$ $\gtrsim$ 3000\,K), two important simplifications occur. First, even their nightside temperatures become too elevated for refractories to condense~\citep[e.g.][]{zhang_phase_2018, wong_exploring_2020, addison_toi-1431bmascara-5b_2021}, meaning that the gas phase should include the entirety of the atmospheric metallic budget. Secondly, the chemistry simplifies, with the temperature being too elevated for molecules to be thermally stable, with even strongly bound CO molecules dissociating past $\sim$4000\,K at sub-millibar pressures (Fig.~\ref{fig:chemistry}, left panel). At 5000\,K, assuming chemical equilibrium, the C and O budget of a planetary atmosphere is almost entirely dominated by neutral atoms at the millibar to microbar pressure levels probed by transit observations, with the temperature not yet high enough for these to respectively ionise into C$^{+}$ and O$^{+}$ in large amounts (Fig.~\ref{fig:chemistry}, right panel). In this extreme temperature regime, the relative proportion of gas-phase atomic C and O should therefore directly reflect the C/O ratio of the bulk envelope. Meanwhile, molecules such as CO and OH can still be stable in higher-pressure, cooler regions of the atmosphere, below the strongly irradiated inverted upper atmosphere~\citep[e.g.][]{yang_detection_2025}.

With an equilibrium temperature of approximately 4000\,K, KELT-9b is the hottest currently known transiting exoplanet orbiting a main-sequence star~\citep{gaudi_giant_2017}. Driven by the intense irradiation it receives from its host, the upper atmosphere of KELT-9b is heated well above the equilibrium temperature on both its dayside~\citep{mansfield_evidence_2020, wong_exploring_2020, kasper_confirmation_2021} and high-altitude terminator regions~\citep{fossati_data-driven_2020, turner_detection_2020, wyttenbach_mass-loss_2020}. At such elevated temperatures, opacity sources at photospheric pressures are likely to be almost entirely from atoms and ions rather than molecules, similar to early K-type stars~\citep{gaudi_giant_2017}. Recently, neutral atomic O was detected in the transmission spectrum of KELT-9b~\citep{borsa_high-resolution_2022} using the CARMENES spectrograph. In this work, we aim to simultaneously detect neutral C and O from transit observations of KELT-9b taken with the MAROON-X spectrograph to measure its atmospheric C/O ratio using atomic species rather than molecules as would typically be done for any other (colder) planet.

\section{Methods}

\subsection{Observations}\label{subsec:obs}

We used the high resolution ($R = 85\,000$) red optical (500 -- 920\,nm) MAROON-X spectrograph~\citep{seifahrt_development_2016, seifahrt_maroon-x_2018, seifahrt_-sky_2020, seifahrt_maroon-x_2022} installed on Gemini-North to observe two transit time series of the ultra-hot Jupiter KELT-9b ($R_p = 1.891_{-0.053}^{+0.061}$\,$R_{\rm Jup}$, $M_p = 2.88\pm0.84$\,$M_{\rm Jup}$, $P = 1.4811235(11)$\,days, $T_0$ $(\mathrm{BJD}_{\mathrm{TDB}}) = 2458711.58627_{-0.000024}^{+0.000025}$, and $T_{\mathrm{eq}} = 4050\pm150$\,K;~\citealt{gaudi_giant_2017, wong_exploring_2020}). The first transit was observed on 2020 May 23, comprising 59 exposures taken over 4.4 hours.  The second transit was observed on 2020 May 26, comprising 65 exposures taken over 4.9 hours.  Both transit time series began slightly after ingress, and included $\sim$1--1.5\,h of baseline after egress. Observing conditions were relatively stable, with only some intermittent clouds causing temporary drops in signal-to-noise during the second night. 

The MAROON-X spectrograph consists of a blue and a red detector; the readout time for the latter is about 40\,s longer.  Exposure times were therefore set to 200\,s for the blue detector and 160\,s for the red detector to maximise photon collection efficiency while maintaining the same cadence. As this resulted in different midpoint times for the exposures from each detector, the time series obtained from each are treated separately in our analysis. The data were optimally extracted using the default MAROON-X reduction pipeline~\citep[see][]{seifahrt_-sky_2020}, providing vacuum wavelength-calibrated spectra (in the telluric rest frame) for each spectral order.

\begin{figure}
    \includegraphics[width=\linewidth]{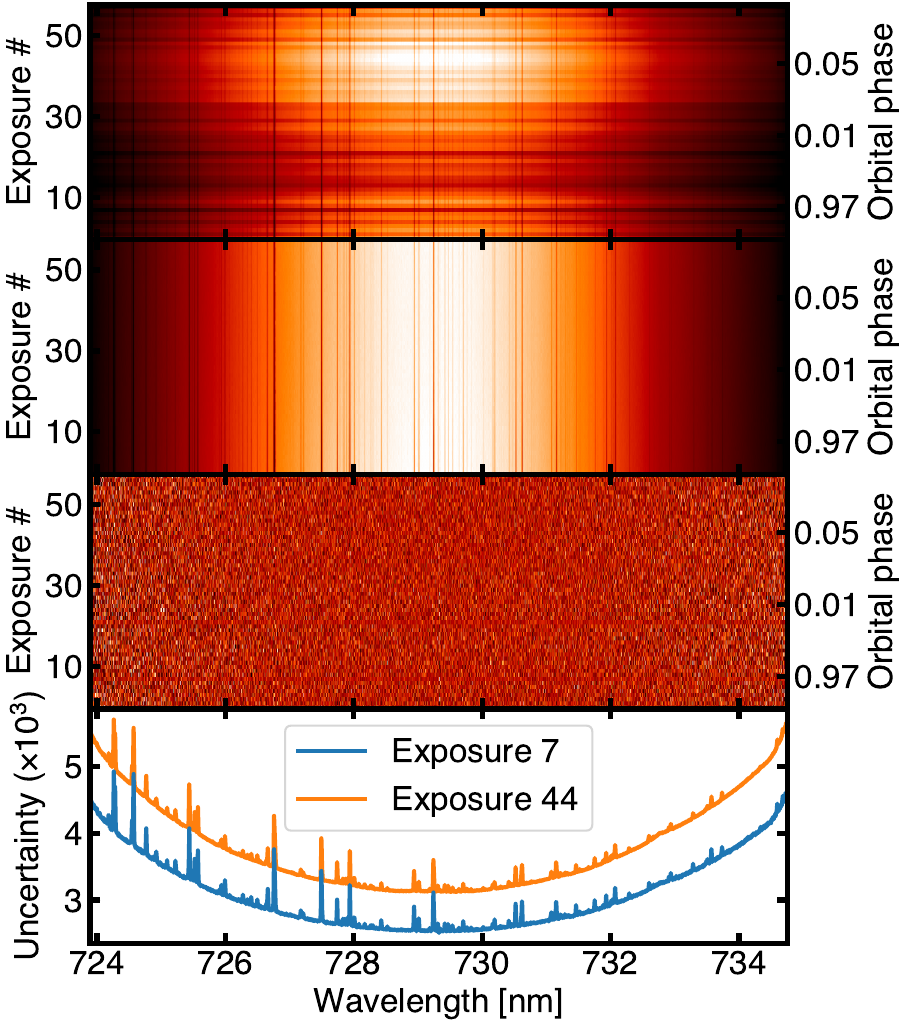}
    \centering
    \caption[]{Data detrending and noise model example.
    \textit{Top panel}: Extracted spectra for a single MAROON-X order centred at 729\,nm for the 4.4-hour transit time series obtained on 2020 May 23. Sharper dark vertical lines depict telluric absorption features, while stellar lines are broader due to KELT-9 being a fast rotator.
     \textit{Second panel}: Data after normalisation and continuum alignment. 
     \textit{Third panel}: Post detrending, where the process removes stellar and telluric features to yield a cleaned product that is used as input for the cross-correlation and retrieval analyses. 
    \textit{Bottom panel}: Example adopted uncertainty per pixel for the lowest (blue) and highest (orange) S/N exposures. Here the rise in noise at the order edges and where telluric lines are present reflects the lower amount of flux received at those wavelengths. Overall, the random nature of the noise without any distinguishable features post detrending indicates the lack of any strong remaining telluric and stellar residuals or uncorrected systematics remaining in the data. 
}
    \label{fig:reduction}
\end{figure}

\subsection{Data detrending}\label{subsec:detrending}
We analysed the data using a similar methodology as in \cite{pelletier_vanadium_2023} and \cite{sun_hat-p-70b_2026}. In brief, we first organised each detector of the two observed MAROON-X transit time series into N$_\mathrm{exposure}$ $\times$ N$_\mathrm{order}$ $\times$ N$_\mathrm{pixel}$ dimensional data cubes (Fig.~\ref{fig:reduction}, top panel). We then Doppler-shifted and cubic-spline-interpolated the spectra to be aligned in the rest frame of the star, correcting for both the motion of the Earth around the barycentre and for the reflex motion caused by KELT-9b on its stellar host. This was done so that star lines remain at a fixed wavelength position throughout a time series. Notably, moving to the stellar rest frame comes at the trade-off of telluric line positions no longer being fixed.  While detrending the data in the telluric rest frame is typically optimal for near-infrared observations with significant contribution from Earth's atmosphere, moving to the stellar rest frame is generally more beneficial for optical data where stellar lines are the main source of contamination~\citep{pelletier_crires_2025, vaulato_hydride_2025}. We find that even in the case of KELT-9, which is a fast rotator with broad spectral features, shifting the data to the stellar rest frame before detrending reduces residuals in the cross-correlation results. 

During planetary transits, observed disc-integrated stellar lines will be distorted due to the Rossiter-McLaughlin (RM) effect~\citep{rossiter_detection_1924, mclaughlin_results_1924}, or `Doppler shadow'. To correct for this, we followed the methodology outlined in \cite{lam_secrets_2024} and used \texttt{StarRotator}~\citep{hoeijmakers_hoeijmakersstarrotator_2024} to model the line distortions caused by KELT-9b obscuring different parts of the projected stellar surface during the transit event.  We modelled the Doppler shadow at the spectral level across the full wavelength range of MAROON-X, assuming a PHOENIX model~\citep{husser_new_2013} as the stellar spectrum ($T_{\mathrm{eff}} = 10{,}000$\,K, $\log g = 4.0$, and [Fe/H] = 0.0).  We then divided the data by this RM model, correcting for these line distortions (e.g. Fig.~\ref{fig:CCF_trail}). We then corrected for bad pixels following \cite{brogi_carbon_2014} before then normalising and continuum-aligning all spectra, order by order, to account for blaze and throughput variations (Fig.~\ref{fig:reduction}, second panel).  This continuum alignment step is done as in \cite{gibson_relative_2022} and consists of first running a box filter (width = 501 pixels) on the data after removing the median spectrum, further smoothing this filter with a Gaussian kernel (standard deviation = 100 pixels), and then subtracting this filter from the data to remove any continuum variations between exposures. We then apply a weighted principal component analysis \citep[wPCA;][]{delchambre_weighted_2015} to further rid the data of remaining stellar and telluric residuals (Fig.~\ref{fig:reduction}, third panel). The median spectrum removal and wPCA were done in magnitude space, with the first five principal components being subtracted out of each spectral time series before returning the data to flux space.

Although five are removed for the main analysis, we tested removing anywhere between three and ten principal components, finding mostly consistent results in all cases. 
However, stronger structured residuals remained visible in the cross-correlation times series when only three or fewer components were removed.  We therefore chose five components applied to all spectral orders as a compromise between cleaning the data to a satisfactory degree and not risking removing too much of the underlying planetary signal.  While this choice is somewhat arbitrary, we specifically avoided over optimising our data detrending procedure as this can, in some scenarios, lead to false positives~\citep[e.g.][]{cabot_robustness_2019, cheverall_robustness_2023}. We use the full spectral coverage of MAROON-X, with the exception of the 760--770\,nm wavelength range dominated by significant telluric absorption that we mask out from the analysis. Spectral channels (pixels) with a variance larger than 4$\sigma$ that of their local order are also masked to avoid outliers that could introduce spurious signals in the cross-correlation analysis. Finally, uncertainties for all data points are estimated using a Poisson noise model following \cite{gibson_detection_2020, gibson_relative_2022} (see our Fig.~\ref{fig:reduction}, bottom panel).  We refer the reader to \cite{pelletier_vanadium_2023} for a more detailed description of each analysis step.

The detrending efficiently removes spectral contributions that are static or quasi-static, effectively removing all stellar and telluric lines that would otherwise dominate the observed spectra. In contrast, planetary lines from KELT-9b's atmosphere during transit rapidly accelerate in pixel (wavelength) space, leaving them largely unaffected by the (weighted) PCA.

\subsection{Atmospheric forward modelling}\label{subsec:fwd}
We generate transmission templates for KELT-9b using the \texttt{SCARLET} framework~\citep{benneke_atmospheric_2012, benneke_how_2013, benneke_strict_2015, benneke_sub-neptune_2019}. The atmosphere is modelled as 100 layers evenly distributed in log pressure between 10$^{2}$ and 10$^{-8}$\,bar. We use the \texttt{FastChem} package~\citep{stock_fastchem_2018, stock_fastchem_2022, kitzmann_fastchem_2024} to calculate chemically consistent abundance profiles. Cross-sections for line absorbers were computed using \texttt{HELIOS-K}~\citep{grimm_helios-k_2015, grimm_helios-k_2021} based on the Vienna Atomic Line Database  (VALD) line lists~\citep{ryabchikova_major_2015}. We also include continuum opacity from H$^{-}$ (bound-free and free-free; \citealt{gray_observation_2021}), collision-induced absorption from H$_2$–H$_2$ and H$_2$–He interactions~\citep{borysow_collision-induced_2002}, and Rayleigh scattering~\citep{benneke_atmospheric_2012}. The radiative transfer is done at a spectral resolution of $R =$ 250,000, with generated transmission spectra then being broadened first with a transit rotating annulus kernel as implemented in \citet[][see their Appendix A]{boucher_co_2023} assuming a tidally locked limb velocity of 6.5\,km\,s$^{-1}$, and then with both a Gaussian kernel at the instrumental resolution of MAROON-X ($R=$ 85,000) and a box kernel to account for the planetary motion during the average exposure time of 180 seconds. 
 
\subsection{Cross-correlation setup}\label{subsec:ccf}
Cross-correlation functions (CCFs) were computed between the detrended data time series' and generated transmission templates over a range of Doppler shifts from $-$500 to $+$500\,km\,s$^{-1}$ in 1\,km\,s$^{-1}$ steps. Although we corrected for line distortions from the RM effect at the spectral level (Sect.~\ref{subsec:detrending}), this correction is unlikely to be perfect over the full spectral range. Therefore, we still chose to discard all data that overlap in velocity space with the Doppler shadow~\citep[e.g.][]{hoeijmakers_atomic_2018}, applying a 15\,km\,s$^{-1}$ wide mask centred at $-$38\,km\,s$^{-1}$ (Fig.~\ref{fig:CCF_trail}). CCF values were then phase-folded into $K_p - V_{\mathrm{sys}}$ maps by summation after being interpolated from the stellar rest frame to the planetary rest frame for different assumed combinations of the planetary Keplerian ($K_p$) and systemic ($V_{\mathrm{sys}}$) velocities. Species-only models can then be used in the cross-correlation to identify individual atmospheric absorbers, which in the case of KELT-9b, has led to a wealth of species being detected in its transmission spectrum \citep[e.g.][]{hoeijmakers_atomic_2018, hoeijmakers_spectral_2019, yan_extended_2018, borsa_high-resolution_2022, borsato_mantis_2023}. While a cross-correlation analysis is not the primary goal of this study, it is still a good metric to visually confirm that all species of interest are indeed detected in the data and to evaluate the effectiveness of the detrending process at removing stellar and telluric features. 

\subsection{Retrieval prescription}\label{subsec:retrieval}
To further characterise the atmosphere of KELT-9b, we utilised a high-resolution retrieval approach~\citep[e.g.][]{brogi_retrieving_2019}.  We used the likelihood formalism of \cite{gibson_detection_2020, gibson_relative_2022}, and \texttt{emcee}~\citep{foreman-mackey_emcee_2013} as a sampler. For each step in the retrieval, a transmission template (Sect.~\ref{subsec:fwd}) is generated for a given set of parameters and convolved with the rotating slice and instrumental broadening kernels. The model is then projected in time along the transit for a given $K_p$ and $V_{\mathrm{sys}}$. A box filter is then applied to the model time series to account for the line-of-sight acceleration of KELT-9b within the duration of each exposure, inducing an additional 2 -- 3\,km\,s$^{-1}$ blurring for the 160 -- 200\,s exposure times of the MAROON-X data. We then injected this modelled transit time series into a PCA reconstruction of what was removed from the data in Sect.~\ref{subsec:detrending} and used the model filtering algorithm of \cite{gibson_relative_2022} to reproduce any alterations that the data detrending has on the planetary signal~\citep{brogi_retrieving_2019}. This time-projection, blurring, and filtering of the atmosphere model is done separately for each data cube and then used to compute the likelihood in the retrieval. 

\begin{figure}[t!]
    \includegraphics[width=\linewidth]{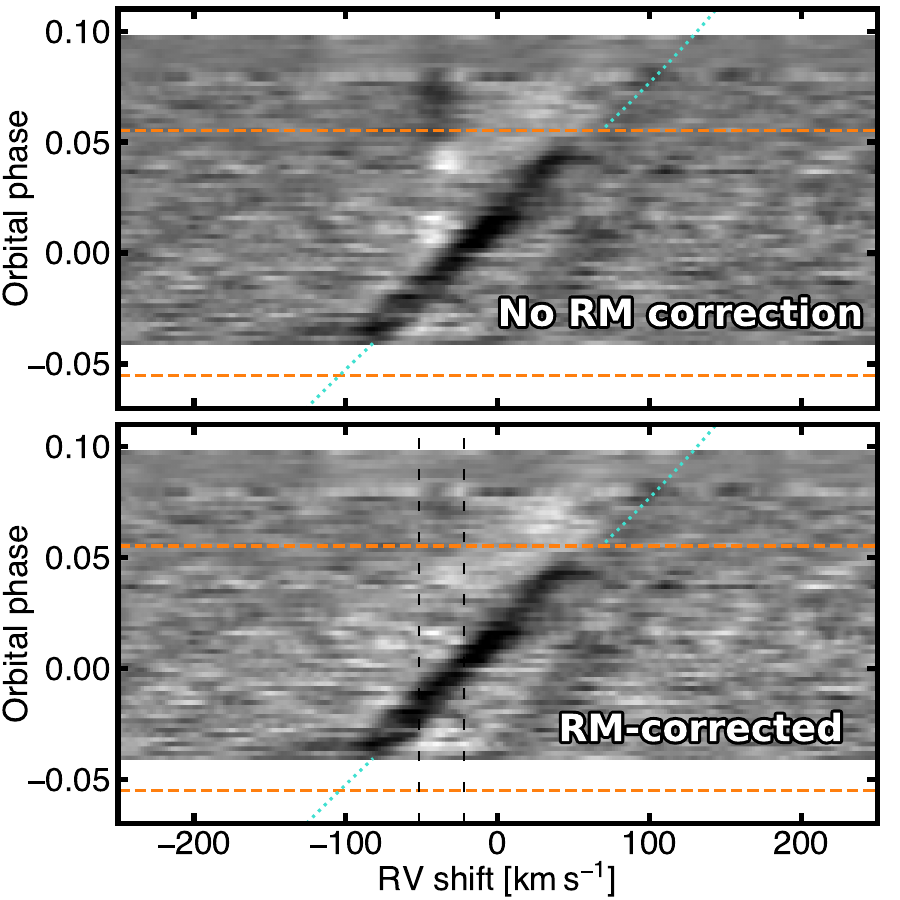}
    \centering
    \caption[]{Orbital trace of KELT-9b's atmosphere during transit and applied RM correction. \textit{Top}: Combined cross-correlation of both detrended MAROON-X transit time series (after wPCA) and a generated model template of KELT-9b. The atmospheric signal from the planet can be seen as a dark trace following its expected orbital motion (dotted black line) over the duration of the transit (dashed orange lines).  Remaining residuals from the imperfectly corrected Doppler shadow can be observed near $-40$\,km\,s$^{-1}$. The transit ingress was missed on both nights of observations.
    \textit{Bottom}: Same but with the \texttt{StarRotator} RM correction now applied. As this correction may not be perfectly accurate, we still discarded all CCF values centred on the Doppler shadow in velocity space (vertical dashed black lines).
    Exposures in which the planetary signal overlaps with the RM trace are not considered in the atmospheric retrieval, to avoid these biasing the results.
    }
    \label{fig:CCF_trail}
\end{figure}

For the likelihood evaluation, we excluded the exposures where the planetary trace overlaps in velocity space with the Doppler shadow (Fig.~\ref{fig:CCF_trail}, masked region). We note that post-PCA residuals from distortions to the stellar lines caused by the RM effect can nevertheless remain present in the data, with the applied mask only ensuring no direct correlation between these and the modelled planet atmosphere. To avoid biases from non-local thermodynamic equilibrium (NLTE) effects, we also masked out all spectral lines for which these are important, leaving only spectral regions where assuming local thermodynamic equilibrium (LTE) is accurate to within a few percent (see Sect.~\ref{sec:caveats}).

\begin{figure*}[t!]
    \includegraphics[width=\linewidth]{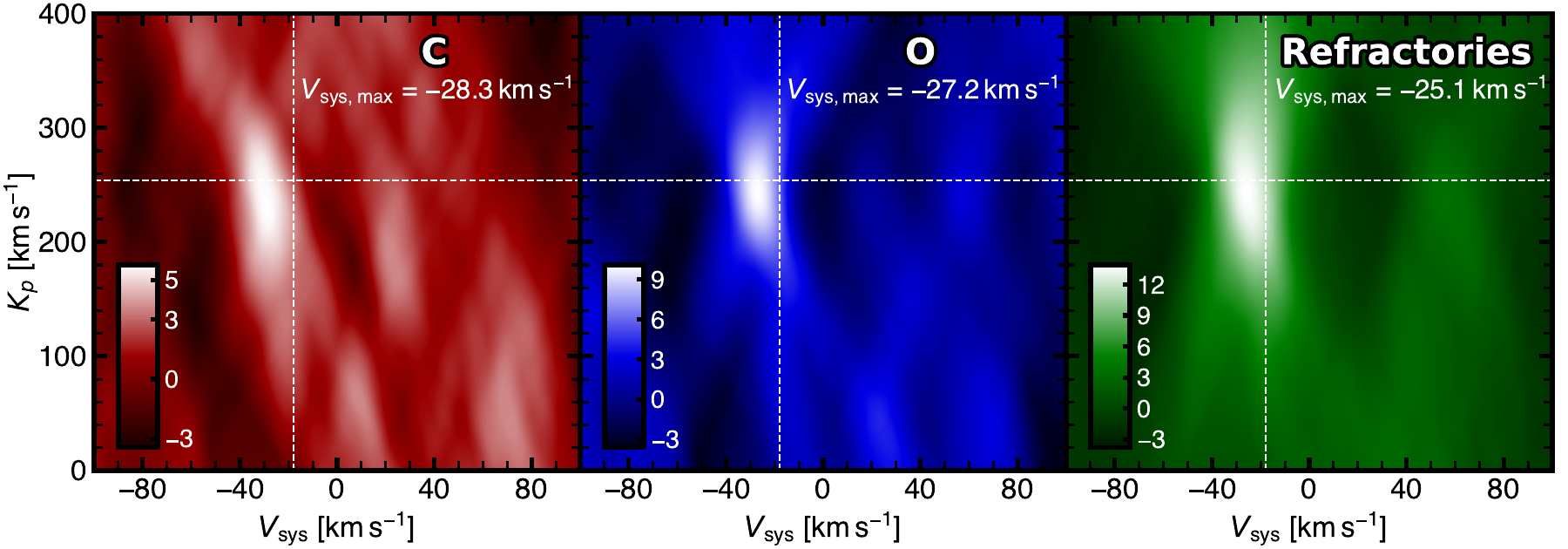}
    \centering
    \caption[]{Cross-correlation signals of neutral atomic C and O and refractories in the atmosphere of KELT-9b. Each panel shows the 2D phase-folded cross-correlation signal-to-noise velocity map of a species, or group of species, relative to the known $K_p$ and $V_{\mathrm{sys}}$ (dashed white lines).  From the MAROON-X data, we recover signals for C, O, and refractory elements (Na, Mg, Ca, Ti, Ti$^{+}$, V, Cr, Fe, Fe$^{+}$, and Ni combined) near the expected velocity positions.  For reference, the peak systemic velocity of each signal is noted, with C and O slightly more blueshifted compared to the refractory metals. The combination of these detections enables the relative volatile and refractory metallicities as well as the C/O ratio of the atmosphere of KELT-9b to be retrieved.
    }
    \label{fig:CCF}
\end{figure*}

As opposed to using a `free’ retrieval approach, in which the abundance of every species is individually fitted for and assumed to be well mixed (constant-with-altitude), we instead opted for a chemical equilibrium approach. The choice for this is twofold. First, this enables us to include opacity contributions for more species without adding a prohibitively large number of parameters by having their individual abundances be governed by an overall metallicity. Secondly, well-mixed abundance profiles are almost certainly a poor assumption for many important chemical components in the atmosphere of KELT-9b. While C and O do not ionise significantly at millibar levels even at 5000\,K (Fig.~\ref{fig:chemistry}, right panel), this is not the case for refractory transition metals (e.g. Fe, Ti, and Ca) that have more weakly bound outer electron shells and hence ionise much more readily. This is particularly evident given the strong detections of numerous ions (e.g. Fe$^{+}$, Ti$^{+}$, and Ca$^{+}$) in the atmosphere of KELT-9b~\citep{hoeijmakers_atomic_2018, hoeijmakers_spectral_2019, yan_ionized_2019, turner_detection_2020, darpa_KELT9_2024}. In the case where neutral and ionised forms of the same species are simultaneously present in the atmosphere (e.g, Fe and Fe$^{+}$), abundance profiles will vary strongly with altitude. Nevertheless, to still maintain some flexibility within our chemical equilibrium retrieval, we chose to characterise different groups of species using different metallicities.  

In the atmospheric retrieval, we let the C and O abundances vary via a combination of the volatile metallicity (M$_{\mathrm{vol}}$) and the C/O ratio. For the refractories, we bundled several of the strongest absorbers in the MAROON-X bandpass detected in the atmosphere of KELT-9b (Na, Mg, Ca, Ti, Ti$^{+}$, V, Cr, Fe, Fe$^{+}$, and Ni) together in a single refractory metallicity (M$_{\mathrm{ref}}$) within which the relative abundances of these species are assumed to be solar-like~\citep{asplund_chemical_2009}.  The purpose of combining these refractories together is to obtain an estimate of the average enrichment of all refractory elements, rather than from a single element (e.g. Fe).  Notably, $[$M$_{\mathrm{vol}}$/H$]$ and $[$M$_{\mathrm{ref}}$/H$]$ are fitted\footnote{Here the notation $[$M/H$]$ refers to the $\log_{10}$ metallicity relative to solar, with $[$M/H$]$ = 1 meaning an atmosphere that is 10 $\times$ solar.} simultaneously but can vary independently, allowing for non-solar ice-to-rock proportions. While more refractory species have been detected on KELT-9b \citep{hoeijmakers_spectral_2019,borsato_mantis_2023}, for computational reasons we only included the ten most significant absorbers in the MAROON-X bandpass in our analysis.  However, we specifically did not include Ca$^{+}$ in the retrieval analysis as its few very strong spectral features can probe up to exospheric pressures~\citep[e.g.][]{yan_ionized_2019,maguire_high-resolution_2023, prinoth_atlas_2024, simonnin_time-resolved_2025, langeveld_time-resolved_2025}. 

We also fitted for an additional parameter, $\log\alpha_{\mathrm{H}^{-}}$, that can vary the abundances of H$^{-}$ and e$^{-}$ used to calculate the continuum opacity from H$^{-}$ bound-free and free-free absorption. In this case, $\alpha_{\mathrm{H}^{-}}$ acts as a multiplicative factor to the abundance profiles of both H$^{-}$ and e$^{-}$ predicted by \texttt{FastChem} for the given temperature structure and composition (set by M$_{\mathrm{vol}}$, M$_{\mathrm{ref}}$, and the C/O ratio). $\log\alpha_{\mathrm{H}^{-}}$ therefore acts as an indirect measure of the ionisation (with a higher value corresponding to more available hydride atoms and free electrons contributing as continuum opacity sources). While both are considered in our analysis, we note that the bound-free opacity contribution is much larger than the free-free contribution over the MAROON-X bandpass (e.g. \citealt{pelletier_crires_2025}; see their Fig. 1).

For the thermal structure, we fitted for a temperature-pressure (TP) profile parameterised by an irradiation temperature $T_{\mathrm{irr}}$, an internal temperature $T_{\mathrm{int}}$, a mean infrared opacity $\kappa_{\mathrm{IR}}$, and the visible-to-infrared opacity ratio $\gamma$ (\citealt{guillot_radiative_2010}; see their Eq. 27). Of these four parameters, we fixed $T_{\mathrm{int}}$ = 200\,K as it has little impact on the end transmission spectrum and fit the remaining three in the retrieval.  The choice of using a \cite{guillot_radiative_2010} profile is a compromise between assuming a potentially oversimplified isothermal profile and fitting for a more complex structure with additional parameters at a significant computational time cost increase.  

We also fitted for an optically thick grey continuum level at a freely parameterised pressure $P_{\mathrm{cont}}$. Although $P_{\mathrm{cont}}$ acts similarly to the H$^{-}$ bound-free opacity, which is only weakly chromatic over the MAROON-X wavelength range~\citep{pelletier_vanadium_2023}, we nevertheless chose to fit both in the retrieval as they affect the transmission spectrum in a fundamentally different way.  While $P_{\mathrm{cont}}$ effectively crops the bottom of all spectral lines by setting the opacity to infinity below a given pressure level, H$^{-}$ absorption plays a more nuanced role by non-uniformly adding opacity across all atmospheric layers, which alters the wings and cores of spectral features. The inclusion of $P_{\mathrm{cont}}$ ensures our model can accurately match the true continuum level in the atmosphere of KELT-9b, given that not all of the plethora of known opacity sources~\citep[e.g.][]{hoeijmakers_spectral_2019,borsato_mantis_2023} are included in our model.

To account for potential additional line broadening sources due to dynamics and the 3D nature of the atmosphere of KELT-9b~\citep[e.g.][]{wardenier_decomposing_2021, wardenier_modelling_2023}, we also included an extra broadening term in the retrieval that convolves the generated templates with a Gaussian kernel of a given full width at half maximum (FWHM). Finally, we fitted for $K_p$ and $V_{\mathrm{sys}}$ in the retrieval as they can deviate from the expected values in ultra hot Jupiter atmospheres observed in transmission~\citep[e.g.][]{prinoth_titanium_2022, prinoth_time-resolved_2023, kesseli_atomic_2022, pelletier_vanadium_2023, wardenier_phase-resolving_2024}.

In all, our retrieval includes four composition parameters ($[$M$_{\mathrm{vol}}$/H$]$, $[$M$_{\mathrm{ref}}$/H$]$, $\log\alpha_{\mathrm{H}^{-}}$, C/O), three temperature structure parameters ($\log \kappa_{\mathrm{IR}}$, $\log \gamma$, $T_{\mathrm{irr}}$), two velocity parameters ($K_p$, $V_{\mathrm{sys}}$), a grey continuum level ($\log P_\mathrm{cont}$), as well as a broadening parameter (FWHM), for a total of 11 free parameters. We set uniform (or uniform in log) priors for all parameters (Table~\ref{tab:retrieval_params}). As in \cite{pelletier_vanadium_2023}, because high-resolution analyses do not take absolute fluxes into account, we also included the photometric transit depth measured from TESS~\citep{wong_exploring_2020} in the retrieval to only allow physically plausible models consistent with the known photometric $R_p/R_{*}$.

\subsection{Systemic velocity measurement}\label{sect:vsys}
We also used the out-of-transit MAROON-X spectra to measure the stellar systemic velocity ($V_{\mathrm{sys}}$) of KELT-9A from its photospheric lines, using the same order-by-order least-squares deconvolution (LSD) and rotational broadening fitting methodology described in \cite{borsato_sorting_2026}, where this analysis is extended across a multi-spectrograph dataset of KELT-9A. This is challenging for such a rapidly rotating star~\citep[$v \sin i = 111.4 \pm 1.3$\,km\,s$^{-1}$,][]{gaudi_giant_2017}, as the broad and shallow stellar lines make it difficult to determine line centres using simple Gaussian fits. We therefore extracted high-S/N mean line profiles using LSD~\citep[][]{donati_spectropolarimetric_1997, kochukhov_least-squares_2010}. LSD combines many photospheric lines into a single average profile while accounting for their rest wavelengths and relative depths, and has been used for radial-velocity measurements of rapidly rotating stars~\citep{borsa_gaps_2019,pai_asnodkar_kelt-9_2022}. We performed the LSD analysis using the \texttt{SpecpolFlow} software package~\citep{folsom_specpolflow_2025}.

For each observation night, we applied LSD to the out-of-transit exposures order by order. Before deconvolution, each echelle order was continuum-normalised with a flexible polynomial model and the order edges were trimmed. The order-by-order LSD profiles were then combined using inverse-variance weights to construct a final profile for each exposure. The stellar line mask was generated using VALD (\citealt{ryabchikova_major_2015}), assuming $T_{\mathrm{eff}} = 10{,}000$\,K, $\log g = 4.0$, and [Fe/H] = 0.0~\citep{gaudi_giant_2017}. We excluded heavily saturated features including the Balmer series, Ca\,\textsc{ii}\,K, and the Ca\,\textsc{ii} near-infrared triplet — as well as the sodium doublet (to avoid interstellar contamination).  This removal was done by setting the uncertainties of the data points surrounding these lines to infinity for the LSD analysis.  The width of the masked region around each line was determined by visual inspection based on the feature extent, varying from around $\pm$500\,km\,s$^{-1}$ from the line centre, up to nearly $\pm$1000\,km\,s$^{-1}$ for the broadest Balmer series lines. The LSD profiles were computed on a velocity grid spanning $-250$ to $+250\mathrm{\,km\,s^{-1}}$, using the provided flux errors as inverse-variance weights.

Each exposure's LSD profile was fitted with the analytic rotational broadening profile of \cite{gray_observation_2021}, assuming a delta-function intrinsic line profile. We implemented this model with the \texttt{RotBroadProfile} class from \texttt{PyAstronomy}~\citep{czesla_pya_2019}, leaving the amplitude ($A$), projected rotational velocity ($v\sin i$), linear limb-darkening coefficient ($\varepsilon$), and line-centre velocity ($\mu$) as free parameters. The model was fitted to each stellar LSD profile using uniform priors of $A \in [-3,0]$, $v\sin i \in [80,140]$~km\,s$^{-1}$, $\varepsilon \in [0,1]$, and $\mu \in [-35,-15]$~km\,s$^{-1}$. The uncertainty on each fitted radial velocity was estimated by propagating the fitted-parameter uncertainties through the same model.

To combine the measurements within each night, we drew 1000 samples from each exposure-level radial-velocity posterior and combined the sampled measurements using inverse-variance weights. The final nightly systemic velocity was taken as the mean of the resulting distribution.

\section{Results and discussion}

Through a cross-correlation analysis, we find signatures of neutral C, neutral O, and various refractory species in the atmosphere of KELT-9b (Fig.~\ref{fig:CCF}). Our results confirm many detections reported in previous work~\citep[e.g.][]{hoeijmakers_atomic_2018, hoeijmakers_spectral_2019, borsa_high-resolution_2022}, while also adding C to the inventory of elements observed in KELT-9b's transmission spectrum. In this case the signal from refractory species includes the combined contributions from Na, Mg, Ca, Ti, Ti$^{+}$, V, Cr, Fe, Fe$^{+}$, and Ni, with the strongest contribution being from Fe and Fe$^{+}$. Notably, the maxima of the CCF signals are all slightly below the expected value of $K_p = 254_{-7}^{+8}$\,km\,s$^{-1}$~\citep{gaudi_giant_2017}. Differences between the true Keplerian orbital velocity of a planet and that measured from an absorber in its atmosphere can naturally occur in ultra-hot Jupiters, as their inflated daysides can create a geometrical asymmetry that results in an apparent reduction in the $K_p$ values measured from transit observations (e.g. \citealt{prinoth_titanium_2022}; see their Fig. 2). Meanwhile, C and O are slightly blueshifted relative to the refractories, potentially indicating that the signals from these species might originate from different pressure levels or trace different dynamics. 

\begin{figure*}[t!]
    \includegraphics[width=\linewidth]{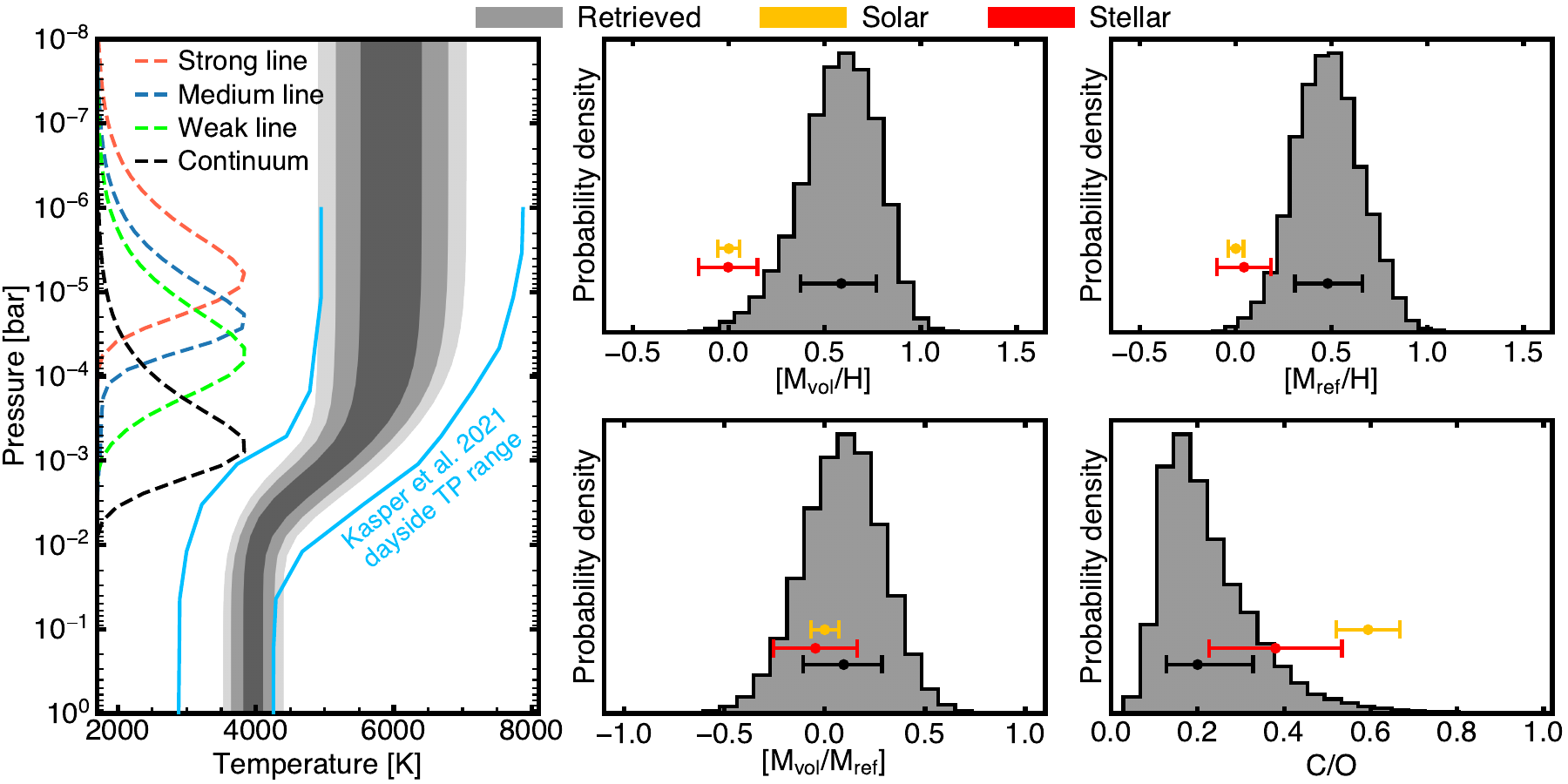}
    \centering
    \caption[]{Overview of the atmosphere retrieval results. 
    \textit{Left}: Retrieved vertical temperature structure (grey; 67\%, 95\%, and 99.7\% confidence intervals) compared to the TP profile range constrained by \cite{kasper_confirmation_2021} from three dayside visits of KELT-9b, also using MAROON-X. Also shown are contribution functions (dashed lines) depicting the pressures probed by spectral lines of various depths and by the continuum set by H$^{-}$.  Note that the strong lines probing the lowest pressures are masked out from our retrieval analysis.
    \textit{Top middle}: Retrieved marginalised posterior distribution for the volatile (carbon and oxygen) metallicity of KELT-9b's atmosphere, compared to the solar~\citep[][orange]{asplund_chemical_2021} and stellar~\citep[][red]{kama_kelt-9_2023} values. 
    \textit{Top right}: Same but for the refractory metallicity. 
    \textit{Bottom middle}: Inferred log volatile-to-refractory metallicity ratio ([M$_\mathrm{vol}$/H] $-$ [M$_\mathrm{ref}$/H]), showing a marginally elevated volatile level of enrichment. 
    \textit{Bottom right}: Retrieved C/O ratio, revealing the atmosphere of KELT-9b to be slightly substellar and significantly sub-solar. Overall, the atmosphere of KELT-9b has a strong thermal inversion and an oxygen-rich (low C/O), super-solar composition.
    }
    \label{fig:TP_abun}
\end{figure*}

From our retrieval analysis, we find that the transmission spectrum of KELT-9b is best described by a TP profile containing a strong thermal inversion (Fig.~\ref{fig:TP_abun}, left panel). The vertical thermal structure is broadly consistent with the dayside TP profile of KELT-9b retrieved by \cite{kasper_confirmation_2021} from MAROON-X thermal emission observations. The temperature notably exceeds 5000\,K at sub-millibar pressures, as predicted by some self-consistent 1D radiative-convective-thermodynamic equilibrium models~\citep[e.g.][]{lothringer_extremely_2018, fossati_non-local_2021}, highlighting the extreme irradiation received by this planet compared to even other ultra-hot Jupiters~\citep{gaudi_giant_2017}. While our retrieved TP profile is somewhat colder than the self-consistent models of KELT-9b from \cite{fossati_non-local_2021} (see our Fig.~\ref{fig:retrieval_compare}), we note that those TP profiles were computed at the substellar point and hence are expected to be hotter than the limb temperatures probed by our observations.

\subsection{Retrieved atmospheric composition} \label{subsec:composition}

For the atmospheric composition of KELT-9b, we measure a $\log_{10}$ volatile metallicity relative to solar of $[$M$_{\mathrm{vol}}$/H$]$ = $0.59_{-0.21}^{+0.19}$, which is elevated relative to the stellar values~\citep[{$[$C/H$]_{\mathrm{stellar}} = -0.13\pm0.08$, $[$O/H$]_{\mathrm{stellar}} = 0.06\pm0.13$,}][]{kama_kelt-9_2023} (see our Fig.~\ref{fig:TP_abun}, top-middle panel).  Similarly for refractories, the metallic enrichment of $[$M$_{\mathrm{ref}}$/H$]$ = $0.60_{-0.17}^{+0.18}$ is enriched relative to what we would expect based on the host star abundances~\citep[{$[$Fe/H$]_{\mathrm{stellar}} = 0.04\pm0.14$, $[$Ca/H$]_{\mathrm{stellar}} = 0.19\pm0.13$, $[$Na/H$]_{\mathrm{stellar}} = 0.12\pm0.20$,}][]{kama_kelt-9_2023} (see our Fig.~\ref{fig:TP_abun}, top-right panel). When combined, the relative proportion of volatiles to refractories (a proxy for the ice-to-rock ratio) is measured to be $[$M$_{\mathrm{vol}}$/M$_{\mathrm{ref}}]$ = $0.09_{-0.20}^{+0.19}$ ($1.24_{-0.78}^{+1.93}$ $\times$ solar), consistent with both the Sun and host star (Fig.~\ref{fig:TP_abun}, bottom-middle panel). The C/O ratio of the atmosphere of KELT-9b is measured to be $0.20_{-0.07}^{+0.13}$, which is lower compared to both the host star values of $0.38\pm0.15$~\citep{kama_kelt-9_2023} and the solar value of $0.59\pm0.07$~\citep{asplund_chemical_2021}.

\begin{table}[t!]
    \caption{Atmospheric retrieval parameter prior ranges and inferred values.} 
    \label{tab:retrieval_params}
    \centering
    \def\arraystretch{1.25}
    \begin{tabular}{ccc}
    \hline
    \hline
    Parameter & Prior & Retrieved value \\
    \hline
    \hline
    $[$M$_{\mathrm{vol}}$/H$]$ & $\mathcal{U}(-3, 3)$ & $0.59_{-0.21}^{+0.19}$  \\
    $[$M$_{\mathrm{ref}}$/H$]$ & $\mathcal{U}(-3, 3)$ & $0.60_{-0.17}^{+0.18}$  \\
    $\log\alpha_{\mathrm{H}^{-}}$ & $\mathcal{U}(-3, 3)$ & $-0.30\pm0.23$  \\
    C$/$O & $\mathcal{U}(0, 10)$ & $0.20_{-0.07}^{+0.13}$ \\
    $T_{\mathrm{irr}}$ [K] & $\mathcal{U}(0, 8000)$  & $4543_{-159}^{+146}$ \\
    $\log \kappa_{\mathrm{IR}}$ [m$^2$\,kg$^{-1}$] & $\mathcal{U}(-2, 2)$  &  $-0.22_{-0.21}^{+0.20}$\\
    $\log \gamma$ & $\mathcal{U}(-1, 1)$  & $>0.36$ \\
    $\log P_\mathrm{cont}$ [bar] & $\mathcal{U}(2, -5)$  & $>-2.73$  \\
    $K_p$ [km\,s$^{-1}$] & $\mathcal{U}(200, 300)$ &  $234.5\pm2.8$ \\
    $V_\mathrm{sys}$ [km\,s$^{-1}$] & $\mathcal{U}(-40, 0)$  &  $-25.7\pm0.6$ \\
    FWHM [km\,s$^{-1}$] & $\mathcal{U}(0, 50)$ &  $<14.3$ \\
    \hline
    $[$M$_{\mathrm{vol}}$/M$_{\mathrm{ref}}]$ & -- & $0.09_{-0.20}^{+0.19}$ \\
    \hline
    \multicolumn{3}{l}{\small $^{*}$for unbounded parameters, 3$\sigma$ lower/upper limits are provided.}\\
    \end{tabular}\\ 
\end{table}

If we assume that the measured atmospheric composition is representative of the primordial accreted envelope, the substellar C/O ratio value combined with the enhanced metallicity could indicate a formation history involving a significant accretion of O-rich solids. Such a scenario would naturally occur, for example, if KELT-9b accreted large amounts of pebbles and/or planetesimals in regions of the protoplanetary disc where water ice was condensed while more volatile carbon preferentially remained in gaseous form~\citep[e.g.][]{oberg_effects_2011,madhusudhan_toward_2014, espinoza_metal_2017, cridland_connecting_2019}. This would likely imply that KELT-9b formed at a much wider separation than its present day orbital position and subsequently underwent inward migration. Such a scenario could be plausible given the misaligned, near-polar orbit of KELT-9b~\citep{gaudi_giant_2017, borsa_gaps_2019}, which may have resulted from a dynamical past interaction with a stellar flyby or another planet~\citep{kozai_secular_1962, nagasawa_formation_2008}, although this is difficult to know as a certainty.

We note that our relatively low measured C/O ratio for KELT-9b stands in contrast to the generally super-solar or consistent with solar C/O ratios measured in other ultra-hot Jupiters such as WASP-76b, WASP-121b, MASCARA-1b, WASP-178b, and WASP-189b that are still in the molecule-dominated (T$_\mathrm{eq} \simeq $ 2000 -- 2800\,K) temperature regime~\citep[e.g.][]{ramkumar_high-resolution_2023, ramkumar_new_2025, cont_exploring_2024, hood_atmospherix_2024, weiner_mansfield_metallicity_2024, smith_roasting_2024, lesjak_retrieving_2025, pelletier_crires_2025, pelletier_enriched_2026, sanchez_stellar_2026}.  However, given the limited sample size and their more than 1000\,K lower equilibrium temperature, it is uncertain whether KELT-9b is more oxygen-rich relative to the overall population of ultra-hot Jupiters, or rather if this is a bias from the different inference methods employed (i.e. inferring the C/O ratio from atomic species rather than molecules). Meanwhile, our measured metal enrichment is consistent with previous studies~\citep[e.g.][]{pino_neutral_2020}, although we note that absolute abundances derived from transit observations are correlated with the continuum opacity level as well as the temperature (via the scale height), and hence can be sensitive to model assumptions. However, the ratios of carbon to oxygen or volatiles to refractories should be more robust~\citep[e.g.][]{benneke_atmospheric_2012, brogi_retrieving_2019, gandhi_retrieval_2023}.

For the combination of the continuum opacity sources, we retrieve a value of $-0.30\pm0.23$ for the $\log$ H$^{-}$ abundance multiplicative factor $\log\alpha_{\mathrm{H}^{-}}$ and a 3$\sigma$ upper limit of $-2.73$ for $\log P_\mathrm{cont}$ (Table~\ref{tab:retrieval_params}). This places the continuum near the millibar level, and the strength of H$^{-}$ bound-free + free-free absorption consistent within 1.5$\sigma$ with that expected from chemical equilibrium predictions at the temperatures retrieved. We note that since $\log\alpha_{\mathrm{H}^{-}}$ and $\log P_\mathrm{cont}$ are strongly correlated due to their similar effect on the modelled transmission spectra, the two parameters cannot be well constrained simultaneously in the retrieval (Fig.~\ref{fig:corner}). The presence of a strong continuum at optical wavelengths (from either H$^{-}$ or other opacity sources) is consistent with the H$^{-}$ bound-free signature reported by \cite{jacobs_strong_2022}.

\subsection{Orbital parameters} \label{subsec:KpVsys}

For the orbital parameters, we recover a value of $K_p = 234.5\pm2.8$\,km\,s$^{-1}$, which is consistent with the previously reported values of $K_p = 234.24 \pm 0.90$\,km\,s$^{-1}$~\citep{hoeijmakers_spectral_2019} and $K_p = 239.07_{-5.79}^{+5.83}$\,km\,s$^{-1}$~\citep{pai_asnodkar_variable_2022} measured from other transmission datasets of KELT-9b probing metals. Differences in these values are not necessarily surprising, as they will ultimately depend on the species included in the cross-correlation template given that they can have varying line strengths and thus can probe different atmospheric layers. In our case, the value for $K_p$ is from the combined signal of all species included in the retrieval, which may not be the same as what we would get if only considering single species templates. For example, \cite{yan_extended_2018} measure a significantly higher $K_p = 268.7_{-6.4}^{+6.2}$\,km\,s$^{-1}$ for KELT-9b from the Balmer H$\alpha$ line probing the extended hydrogen envelope. An important distinction to make here is that the `$K_p$' measured from an atmospheric signature is not necessarily equal to the true Keplerian orbital velocity of KELT-9b. Rather, this $K_p$ also encompasses information about the dynamics, non-uniformness, and 3D nature of the atmosphere~\citep{wardenier_decomposing_2021, wardenier_modelling_2023, wardenier_phase-resolving_2024, savel_no_2022, beltz_magnetic_2022}, and can differ for different chemical species probing different regions of the atmosphere (e.g. \citealt{borsato_mantis_2023}; see their Fig. 6).  

When compared to the known systemic velocity of the star-planet system, the $V_{\mathrm{sys}}$ measured from the planetary atmosphere in transmission can trace the speed of day-to-night winds~\citep[e.g.][]{snellen_orbital_2010}. Here we retrieve a value of $V_\mathrm{sys} = -25.7\pm0.6$\,km\,s$^{-1}$ for KELT-9b, which is blueshifted relative to the MAROON-X stellar systemic velocity of $-18.01 \pm 0.35$\,km\,s$^{-1}$ measured from the host star (see Appendix~\ref{appendix:RV} and Fig.~\ref{fig:rv_maroonx}). However, we note that KELT-9 is a fast rotating A type star~\citep[$v \sin i = 111.4 \pm 1.3$\,km\,s$^{-1}$,][]{gaudi_giant_2017}, making an accurate and precise inference of its absolute systemic velocity difficult, as evidenced by the contrasting estimates reported in the literature (e.g. \citealt{ridden-harper_KELT9_2023}; see their Table 8) even when using the same instrument (e.g. \citealt{pai_asnodkar_kelt-9_2022}; see their Table 3).  Given this uncertainty on the measured velocity of the KELT-9 system, we opted not to over-interpret our measured $V_{\mathrm{sys}}$ from the planetary signal, instead noting that the implied planet-frame velocity offset is $\Delta V_{\mathrm{offset}} = -7.7 \pm 0.6$\,km\,s$^{-1}$, consistent with a net blueshift from day-to-night winds.  While such strong winds are consistent with some previous estimates in the literature~\citep[e.g.][]{pino_gaps_2022, zhang_extreme_2026}, others can range from being consistent with no winds~\citep{hoeijmakers_spectral_2019, cauley_atmospheric_2019}, or even stronger winds of order $\sim$10-12\,km\,s$^{-1}$ in the upper atmosphere~\citep{pai_asnodkar_kelt-9_2022}.  However, we again stress that a direct comparison of the day-to-night wind speeds derived from transmission spectroscopy by different studies should not necessarily be compared directly as these will depend both on the assumed reference $V_{\mathrm{sys}}$ and the ephemeris~\citep{pai_asnodkar_variable_2022, smith_combined_2024, pelletier_breaking_2026}. Further complicating this picture, the winds on KELT-9b may also be variable of the order of $\sim$5–8\,km\,s$^{-1}$~\citep{pai_asnodkar_variable_2022}.

Finally, the FWHM of the Gaussian kernel applied to the final transmission spectrum in addition to the rotational, instrumental, and exposure broadening kernels is retrieved as an upper limit (3$\sigma$) to be below $12.6$\,km\,s$^{-1}$. Its effect being relatively small indicates that the extent of the atmospheric signal in velocity space is largely explained by the included modelled broadening kernels (Fig.~\ref{fig:spec_kernel}, right panel), without the need for significant additional sources from, for example, dynamical or 3D effects.

\section{Modelling limitations and caveats}\label{sec:caveats}

While determining the C/O ratio of a planetary atmosphere via atomic species is a novel approach, its applicability is limited and primarily relevant only to temperature regimes more typical of stellar photospheres (T $\gtrsim$ 4000\,K; Fig.~\ref{fig:chemistry}). In particular, there are some caveats regarding the application of our modelling framework to such extreme atmospheres, which we discuss here.

For oxygen, we find that the CCF detection is in part driven by the O triplet at 777.4\,nm, which is known to be particularly sensitive to NLTE effects~\citep{kiselman_777_1993, steffen_photospheric_2015, amarsi_inelastic_2018, bergemann_solar_2021}. Indeed, \cite{borsa_high-resolution_2022} found evidence of NLTE effects in the transmission spectrum of KELT-9b from excess absorption of this triplet. Other works have also found that NLTE models are necessary for matching various deep spectral features in transit observations of KELT-9b and other ultra-hot Jupiters~\citep{fossati_non-local_2021, fossati_gaps_2023, fossati_non-local_2025, darpa_KELT9_2024, stangret_gaps_2024, baldwin_high-resolution_2026}. Given that our modelling framework assumes LTE, it is inadequate for fitting deep spectral lines that are strongly affected by NLTE effects, which could result in significant biases of any inferred atmospheric properties.

\begin{figure}[t!]
    \includegraphics[width=\linewidth]{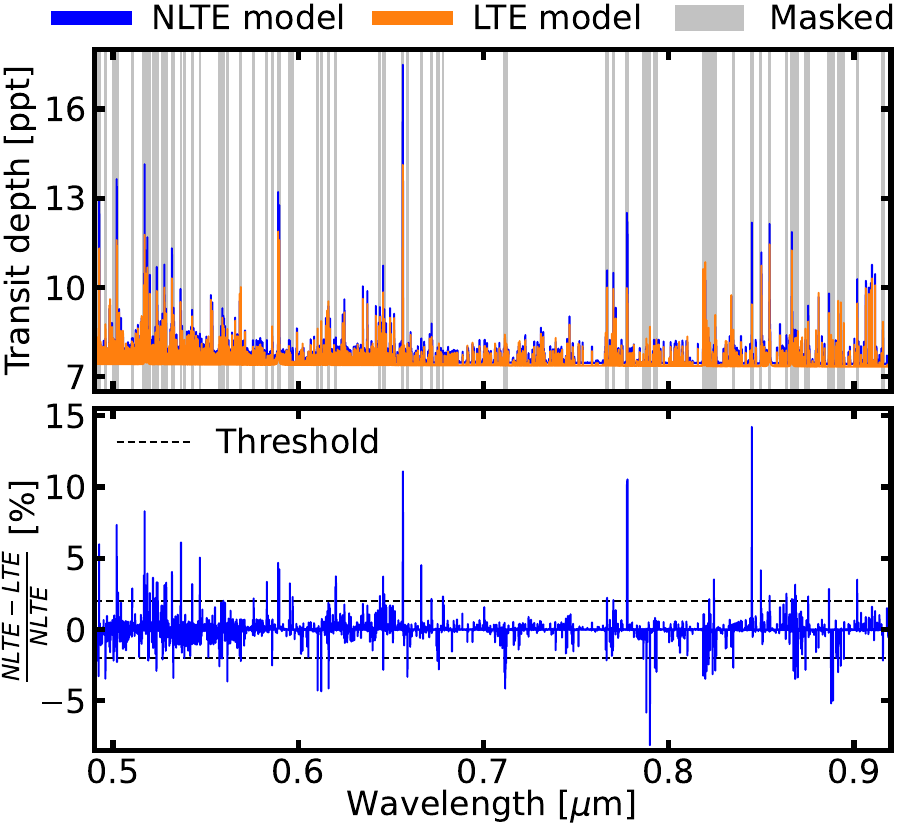}
    \centering
    \caption[]{Identification and masking of spectral lines strongly affected by NLTE effects. 
    \textit{Top}: Comparison between the modelled LTE (orange) and NLTE (blue) transmission spectra of KELT-9b from \cite{fossati_non-local_2021}. The grey shading represents masked wavelength regions where NLTE effects are important. 
    \textit{Bottom}: Percentage difference between the LTE and NLTE transmission spectra. All lines ($\pm$100\,km\,s$^{-1}$) that deviate by more than 2\% (dashed black lines) are masked (grey shading in the top panel) from the retrieval analysis to minimise biases from our models that assume an atmosphere in LTE.
    }
    \label{fig:mask}
\end{figure}

To circumvent this issue, we compared the LTE and NLTE models of KELT-9b from \cite{fossati_non-local_2021} in order to mask out from our data all lines for which the difference in predicted transit depth is greater than 2\%, including a $\pm$100\,km\,s$^{-1}$ half-width to cover the radial velocity extent of KELT-9b during transit. This resulted in 13.5\% (8.8\%) of the MAROON-X blue (red) channel wavelength range, including all of the deepest spectral features, being masked out from our retrieval analysis (Fig.~\ref{fig:mask}). Maintaining only wavelength regions that do not deviate by more than 2\% from LTE predictions should mean that our models can adequately fit all observed spectral features.  However, it may be that even for remaining lines, minor NLTE effects could still bias our results.  We tested only keeping wavelengths compatible within 1\% of LTE, but this resulted in too much data being masked and our retrievals no longer converging. We compare how our results change with and without the application of this NLTE line mask in Fig.~\ref{fig:retrieval_compare}.

The retrieved temperature structure also extends to extremely elevated temperatures (T $>$ 6000\,K) at sub millibar pressures. This notably transcends the limitations of \texttt{FastChem}, which only computes equilibrium chemistry for temperatures up to 6000\,K.  While we allowed our parameter space to include temperatures above this limit, the chemistry was still calculated assuming T = 6000\,K for all TP profile layers that go above this threshold. As C ionises more easily than O (Fig.~\ref{fig:chemistry}), predicting the chemical equilibrium composition while underestimating the true temperature could cause the relative proportion of neutral C to neutral O to be slightly overestimated by our model. This, in turn, could cause the C/O ratio to be under-predicted in the hottest, low pressure regions of the atmosphere of KELT-9b. 

Our model also has numerous limitations, which likely prevents us from perfectly matching the observed transmission spectrum of KELT-9b. For one, we are modelling a 3D atmosphere with a 1D hydrostatic plane parallel radiative transfer code.  The mismatch is partly made evident by the shape of the observed CCF signal (Fig.~\ref{fig:CCF}), as well as the asymmetric and variable transit light curve of KELT-9b observed in photometric observations~\citep{cauley_atmospheric_2019, ahlers_kelt-9_2020}.  While 3D general circulation models are likely necessary to perfectly match the velocity extent of the observed signal~\citep[e.g.][]{wardenier_decomposing_2021, wardenier_modelling_2023}, in practice, we only allowed $K_p$ and $V_{\mathrm{sys}}$ to vary and included an added Gaussian broadening parameter on top of a transit rotational kernel.  In our analysis we also assumed that the atmosphere is static in time, using the same atmospheric model to fit the two transits simultaneously.  In reality, the winds in the atmosphere of KELT-9b are potentially variable~\citep{pai_asnodkar_variable_2022}, although its overall climate appears to be stable~\citep{wong_exploring_2020, jones_stable_2022}.  Notably, the two observed MAROON-X transits were taken only three days (two orbital periods) apart, although it is unclear whether this is less than any potential variability timescale.  Similarly, it is possible that KELT-9b has an extended escaping atmosphere~\citep{yan_extended_2018, yan_ionized_2019, turner_detection_2020, wyttenbach_mass-loss_2020, sanchez-lopez_detection_2022, kama_kelt-9_2023, baldwin_high-resolution_2026, zhang_hydrogen_2026}, which could cause line contrasts in the modelled transmission spectra to be underestimated.

The atmosphere of KELT-9b is also extremely rich in detected chemical species (neutral and ionised) spanning the full range of the periodic table~\cite[e.g.][]{hoeijmakers_atomic_2018, hoeijmakers_spectral_2019, ridden-harper_KELT9_2023, borsato_mantis_2023}.  It is currently beyond our capabilities to include the opacity contributions of all atmospheric components.  However, we did not find a significant change in our results when changing from using opacities of the ten strongest refractory absorbers in our model to only including two (Fe and Fe$^{+}$) or four (Fe, Fe$^{+}$, Ti, and Ti$^{+}$) species. Our models also assume that all refractories are governed by a single metallicity, scaling all abundances uniformly assuming a solar-like distribution as implemented in \texttt{FastChem}. However, the relative proportion of elements in the atmosphere of KELT-9b is unlikely to be exactly equal to the solar distribution. While using stellar abundance ratios would likely be a more adequate assumption, these notably do not differ too significantly from those of the Sun~\citep{kama_kelt-9_2023}. Finally, the intense irradiation received by KELT-9b could give rise to photochemical and photo-ionising effects in its atmosphere~\citep[e.g.][]{casewell_multiwaveband_2015, kitzmann_peculiar_2018, fossati_data-driven_2020} that our analysis does not include.

All in all, we acknowledge that the capabilities of our models are unlikely to be adequate to perfectly depict the extreme environment that is the atmosphere of KELT-9b. In particular, it is unclear how any model shortcomings would bias the inferred atmospheric processes.  All results presented in this analysis should therefore be interpreted in the context of these limitations.  Nevertheless, the determination of the C/O ratio via atomic C and O demonstrated here should be robust in concept.  Future investigation of the carbon and oxygen abundances in the atmosphere of KELT-9b with more complex models will be necessary to confirm the substellar C/O ratio measured here.

\section{Conclusions}

We analysed two transits of the ultra-hot Jupiter KELT-9b obtained with the MAROON-X high-resolution spectrograph. Via a cross-correlation analysis we recovered detections of atomic C and O, as well as refractory species. We then used a high-resolution retrieval framework to infer the physical parameters of the atmosphere of KELT-9b, finding a strongly inverted temperature structure and a slightly metal-rich composition. In terms of abundance ratios, we find the atomic C/O ratio of KELT-9b to be $0.20_{-0.07}^{+0.13}$ and the ($\log_{10}$) proportion of volatile to refractory elements relative to solar to be $0.09_{-0.20}^{+0.19}$. If our inferred composition is representative of the overall envelope, our inferred C/O ratio may indicate that KELT-9b accreted a significant portion of O-rich solids during its formation and evolution history.

Although measuring C/O ratios directly from atomic species is only possible for the hottest known planets, this approach mirrors that used for stellar abundance determinations, bringing us closer to directly fitting stellar and planetary spectral lines simultaneously, thereby obtaining more direct planet--star relative abundance comparisons.  As robustly constraining the C/O ratio of KELT-9b is likely impossible to do via molecules alone, except perhaps if probing the nightside, this approach also allows us to relate the volatile composition of KELT-9b to the broader population of cooler ultra-hot Jupiters that can be characterised from H$_2$O, OH, and CO detections.

\begin{acknowledgements}
We thank the anonymous referee for comments that enabled a more robust analysis and overall improved the quality of the manuscript. Based on MAROON-X observations taken at the Gemini North Observatory for program GN-2020A-Q-234 (PI: Bean).
The development of the MAROON-X spectrograph was funded by the David and Lucile Packard Foundation, the Heising-Simons Foundation, the Gemini Observatory, and the University of Chicago. We thank the staff of the Gemini Observatory for their assistance with the commissioning and operation of the instrument. Based on observations obtained at the international Gemini Observatory, a program of NSF NOIRLab, which is managed by the Association of Universities for Research in Astronomy (AURA) under a cooperative agreement with the U.S. National Science Foundation on behalf of the Gemini Observatory partnership: the U.S. National Science Foundation (United States), National Research Council (Canada), Agencia Nacional de Investigaci\'{o}n y Desarrollo (Chile), Ministerio de Ciencia, Tecnolog\'{i}a e Innovaci\'{o}n (Argentina), Minist\'{e}rio da Ci\^{e}ncia, Tecnologia, Inova\c{c}\~{o}es e Comunica\c{c}\~{o}es (Brazil), and Korea Astronomy and Space Science Institute (Republic of Korea). This work was enabled by observations made from the Gemini North telescope, located within the Maunakea Science Reserve and adjacent to the summit of Maunakea. We are grateful for the privilege of observing the Universe from a place that is unique in both its astronomical quality and its cultural significance. 
This project has been carried out within the framework of the National Centre of Competence in Research PlanetS supported by the Swiss National Science Foundation under grant 51NF40\_205606. S.P., D.E.\, M.S.\, V.V.\, and A.R.C.S. acknowledge the financial support of the SNSF. D.E.\ and M.S.\ have also received funding from the Swiss National Science Foundation for project 200021\_200726.
J.L.B.\ acknowledges the financial support from NASA XRP grant 80NSSC19K0293 and NSF grant AST-2307177. 
H.J.H.\ acknowledges the financial support from eSSENCE (grant number eSSENCE@LU 9:3), the Swedish National Research Council (project number 2023-05307), The Crafoord foundation and the Royal Physiographic Society of Lund, through The Fund of the Walter Gyllenberg Foundation. 
B.T.\ acknowledges the financial support from the Wenner-Gren Foundation (WGF2022-0041).
This project has received funding from the European Research Council (ERC) under the European Union’s Horizon 2020 research and innovation programme (project {\sc Four Aces}; grant agreement No 724427). 
A.R.C.S. acknowledges support from Funda\c{c}\~ao para a Ci\^encia e Tecnologia (FCT) and POCH/FSE through the research grants UIDB/04434/2020 and UIDP/04434/2020, and the FCT fellowship 2021.07856.BD. This work has received funding from the European Union (ERC, FIERCE, 101052347).
This publication makes use of The Data \& Analysis Center for Exoplanets (DACE), which is a facility based at the University of Geneva (CH) dedicated to extrasolar planets data visualization, exchange and analysis. DACE is a platform of the Swiss NCCR PlanetS, federating the Swiss expertise in Exoplanet research. 
\end{acknowledgements}

\bibliographystyle{aa}
\bibliography{K9_MX}

\begin{appendix}

\onecolumn
\nolinenumbers

\clearpage

\section{Auxiliary retrieval results}\label{sec:appendix_retrieval}
\begin{figure*}[ht]
    \centering
    \includegraphics[width=0.9\linewidth]{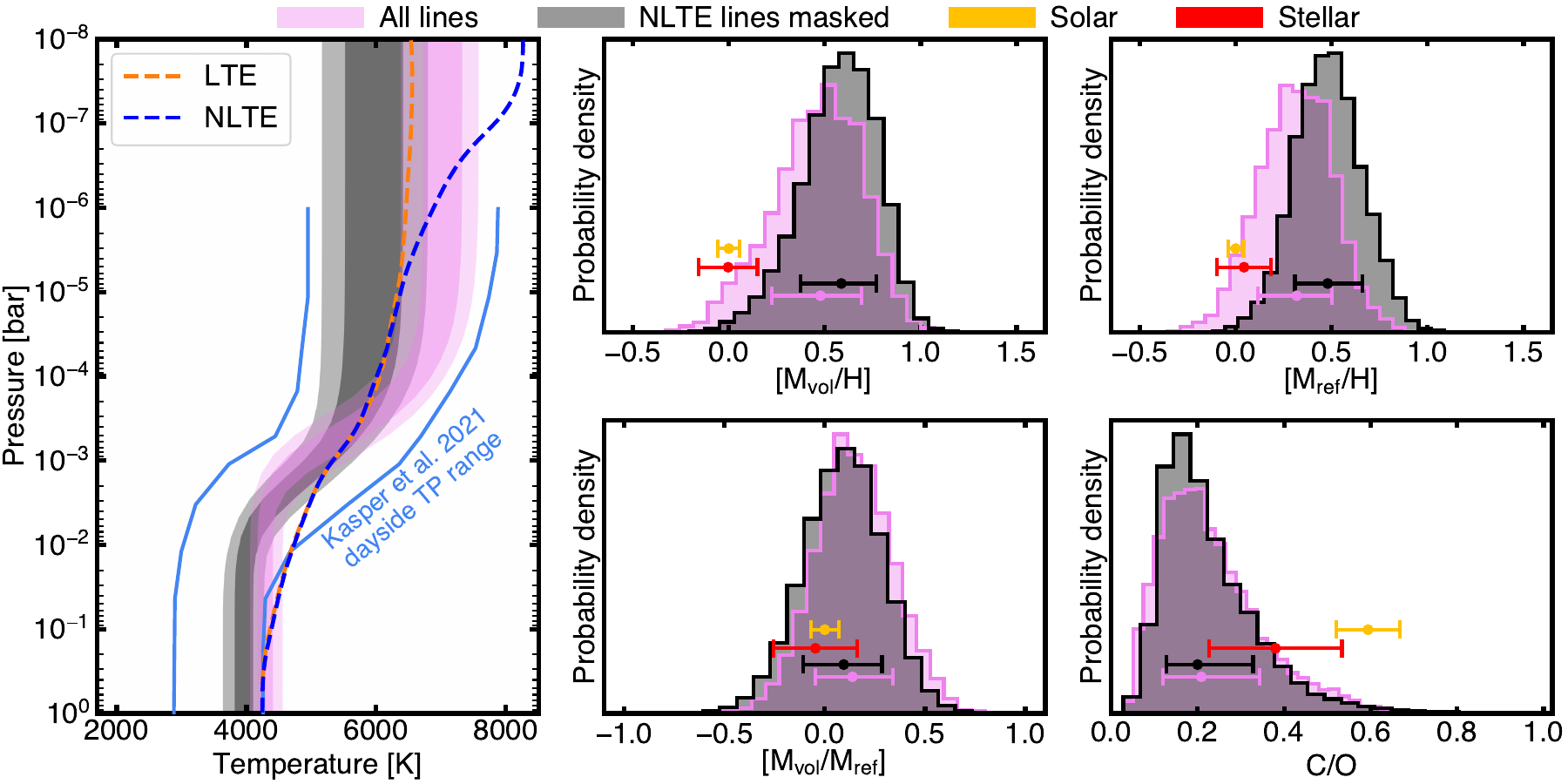}
    \vspace{-3mm}
    \caption{Similar to Fig.~\ref{fig:TP_abun} but now comparing the atmospheric retrievals results without (purple) and with (grey) NLTE-affected lines masked out of the analysis (Fig.~\ref{fig:mask}).  We also compare the retrieved thermal profiles to the self-consistent LTE and NLTE models of KELT-9b from \cite{fossati_non-local_2021}.  Note that only 2$\sigma$ contours are shown on the TP profiles, for visual clarity. Excluding lines affected by NLTE effects by more than 2\% notably results in a slightly cooler retrieved TP profile that is more consistent with the LTE model.
    }
    \label{fig:retrieval_compare}
\end{figure*}

\begin{figure*}[ht]
    \centering
    \includegraphics[width=0.63\linewidth]{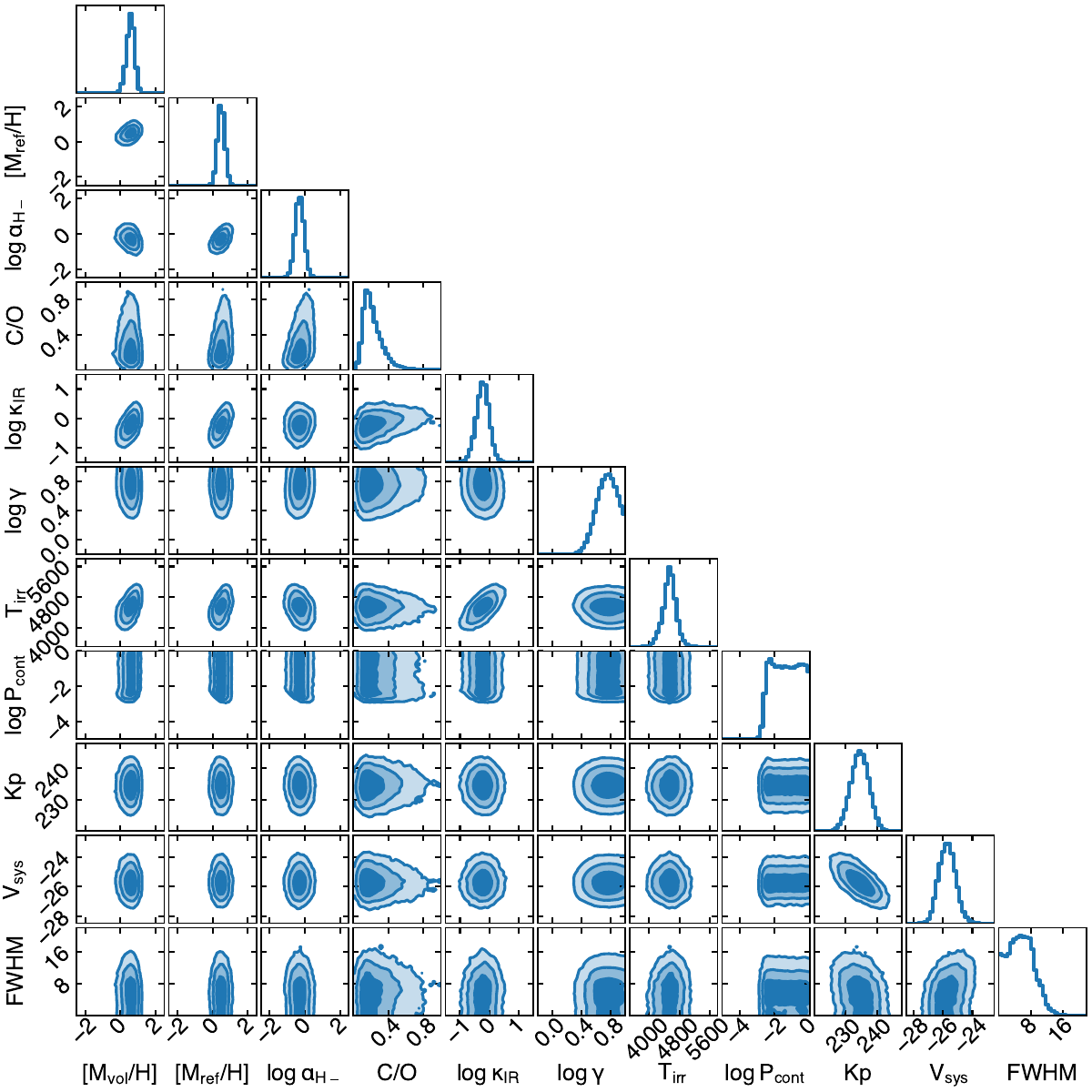}
    \vspace{-3mm}
    \caption{Retrieved constraints on the atmospheric and orbital properties of KELT-9b obtained from two MAROON-X transits.  Shown are the marginalised posterior distributions for the abundance, TP profile parameters, continuum pressure level, velocity, and broadening parameters.  The shaded regions respectively depict the 39.3\%, 86.5\%, and 98.9\% confidence intervals.  
    }
    \label{fig:corner}
\end{figure*}

\begin{figure*}[ht]
    \centering
    \includegraphics[width=0.9\linewidth]{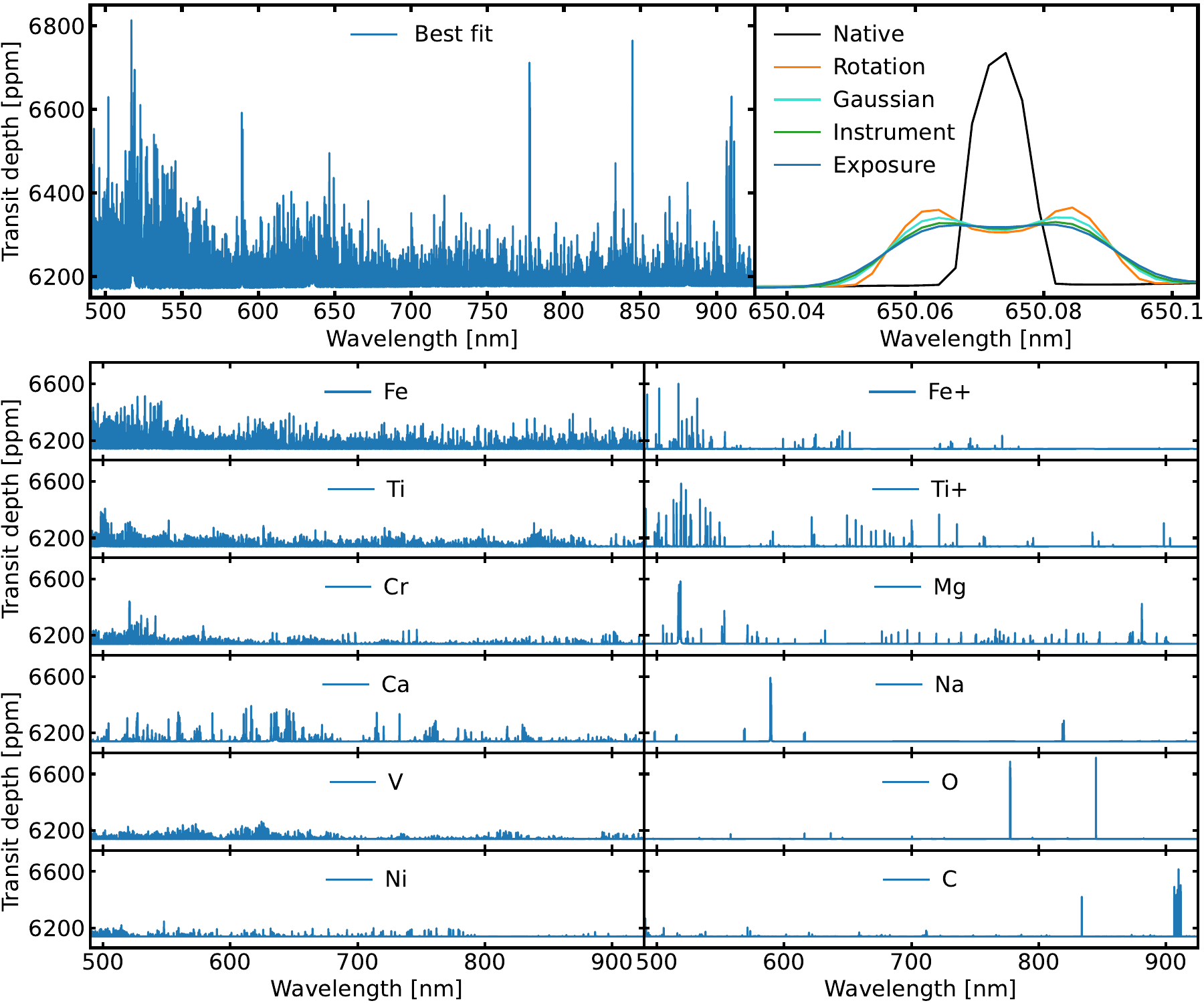}
    \caption{Best-fit spectrum and example line profile.
    \textit{Top left}: Retrieved best-fit model of KELT-9b, with all sources of broadening applied.
    \textit{Top right}: Line profile after the subsequent application of the different included broadening sources. Here an example spectral feature is shown first at the native model resolution (black) and after broadening from the rotational kernel (orange), the extra Gaussian parameter (turquoise), the instrumental resolution (green), and the exposure blurring from the finite integration time of each observed spectrum (blue).
    \textit{Bottom panels}: Individual contributions of the line absorbers included in the model.
    }
    \label{fig:spec_kernel}
\end{figure*}

Here we provide additional retrieval outputs, including a comparison of the retrieval results before and after masking out wavelength regions significantly affected by NLTE effect (Fig.~\ref{fig:retrieval_compare}), the best fit model (Fig.~\ref{fig:spec_kernel}, top-left panel) and how it is affected by each included source of broadening (Fig.~\ref{fig:spec_kernel}, top-right panel), and a corner plot of all retrieved parameters (Fig.~\ref{fig:corner}). 

\vspace{-3mm}

\vspace{-6mm}

\clearpage

\section{Radial velocity distributions for MAROON-X observations}\label{appendix:RV}
\begin{figure*}[ht]
    \centering
    \includegraphics[width=\linewidth]{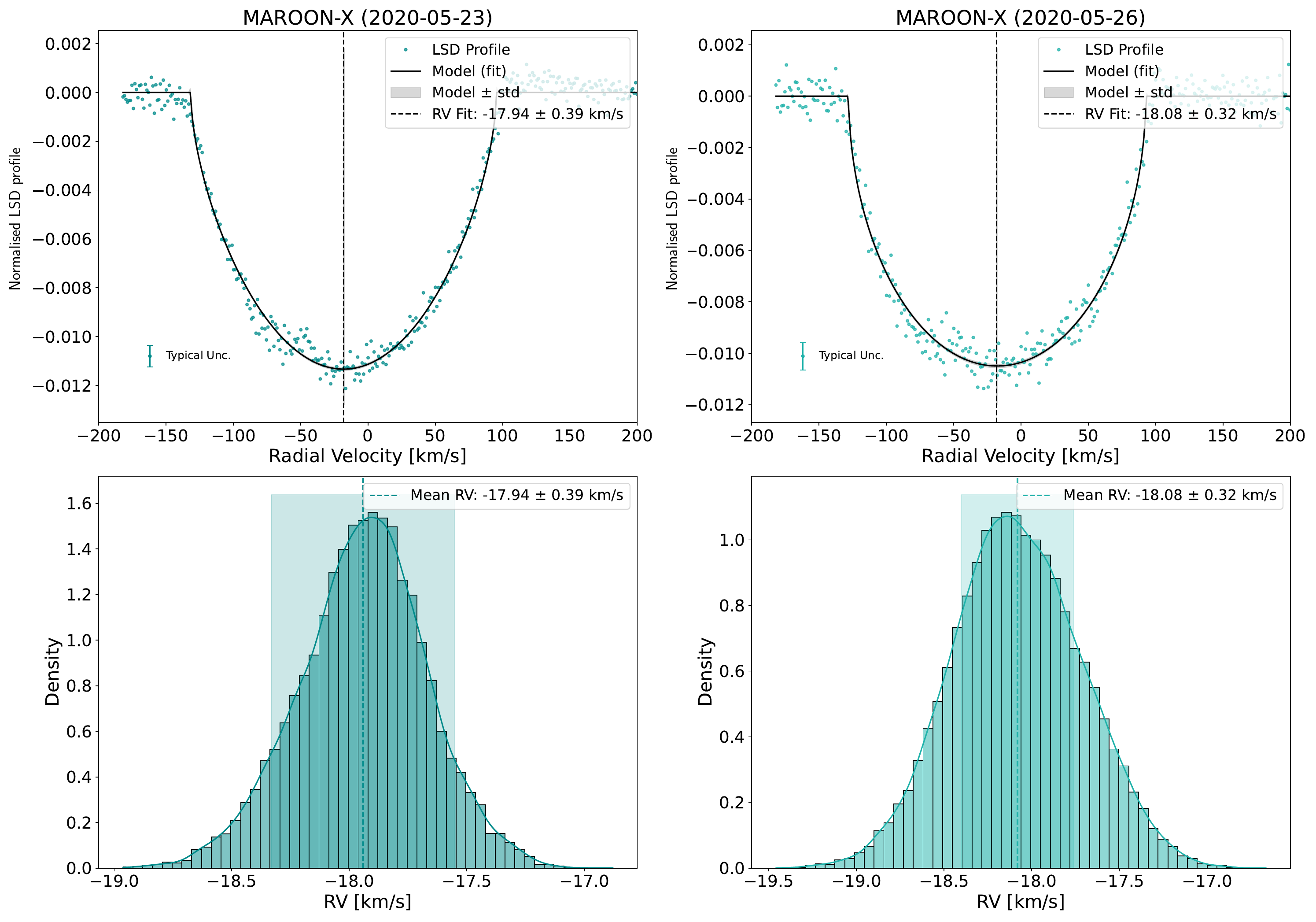}
    \caption{LSD profile fits and radial velocity distributions for the two MAROON-X nights. \textit{Top panels}: Mean least-squares deconvolved stellar profiles from the out-of-transit exposures for each night. Solid lines represent the best-fit rotational broadening models, and grey shaded regions indicate the $1\sigma$ spread across bootstrap realisations. \textit{Bottom panels}: Histograms of the derived stellar radial velocities per exposure. Vertical dashed lines mark the mean recovered velocity, with grey bands showing the $\pm 1\sigma$ interval. These plots illustrate the internal stellar radial velocity scatter and night-to-night stability of the MAROON-X reference velocity.}
    \label{fig:rv_maroonx}
\end{figure*}

The inference of exoplanetary wind measurements from high-resolution transit observations is a differential measurement relative to the reference value of the host star.  Because reported $V_{\mathrm{sys}}$ values in the literature can vary depending on the instrument used, here we measured the reference systemic velocity of the host star for both our observed transits directly from out-of-transit spectra as described in Sect.~\ref{sect:vsys}, following the order-by-order LSD procedure applied across multiple spectrographs by \cite{borsato_sorting_2026}.  We measure $V_{\mathrm{sys}} = -17.94\pm0.39$\,km\,s$^{-1}$ for the observations of 2020 May 23 and $V_{\mathrm{sys}} = -18.08\pm0.32$\,km\,s$^{-1}$ for the transit of 2020 May 26 (Fig.~\ref{fig:rv_maroonx}). Combining the two MAROON-X nights gives $V_{\mathrm{sys}} = -18.01\pm0.35$\,km\,s$^{-1}$. These quoted uncertainties correspond to the standard deviations of the exposure-level distributions.

\end{appendix}
\end{document}